\documentstyle[12pt]{article}
\newcommand{\be}{\begin{equation}}
\newcommand{\ee}{\end{equation}}
\newcommand{\bea}{\begin{eqnarray}}
\newcommand{\eea}{\end{eqnarray}}
\newcommand{\ba}{\begin{array}}
\newcommand{\ea}{\end{array}}

\def\bbox{{\,\lower0.9pt\vbox{\hrule \hbox{\vrule height 0.2 cm
\hskip 0.2 cm \vrule height 0.2 cm}\hrule}\,}}
\newcommand{\dsl}{\pa \kern-0.5em /}

\font\mybb=msbm10 at 10pt
\def\bb#1{\hbox{\mybb#1}}

\def\bR {\bb{R}}
\def\bE {\bb{E}}

\begin{document}


\begin{titlepage}

\vfill

\begin{center}
\baselineskip=16pt
{\Large\bf The Story of M$^*$}
\vskip 0.3cm
{\large {\sl }}
\vskip 10.mm
{\bf ~Paul K. Townsend } \\
\vskip 1cm
{\small
DAMTP, Center for Mathematical Sciences,\\
University of Cambridge, \\
Wilberforce Road,\\
Cambridge CB3 0WA, UK\\
}
\end{center}
\vfill
\par
\begin{center}
{\bf ABSTRACT}
\end{center}
\begin{quote}
The origins of the 11-dimensional supermembrane are recalled, and a
curious property is discussed: the field theory limit of a
supermembrane in a hyper-K\"ahler background is a 3-dimensional
sigma-model with $N=4$ supersymmetry, but the higher-order fermion
interactions of the supermembrane generically break this to $N=3$.
\vskip 3cm

{\it They were learning to draw}, the doormouse went on, \dots {\it
and they drew all manner of things--everything that begins with M}.

{\it Why with an M?} said Alice.

{\it Why not?} said the March Hare.

\vfill
 \hrule width 5.cm
\vskip 2.mm

{\small
\noindent $^*$ Contribution to proceedings of Stephen Hawking's 60th 
birthday conference, {\it The future of theoretical physics and
cosmology}.\\
}
\end{quote}
\end{titlepage}

\setcounter{equation}{0}
\section{Introduction}

Shakespearean actors are traditionally averse to pronouncing the name of
the play `Macbeth', preferring to call it `the Scottish play'. Presumably
it was only distaste for cryptic abbreviations that prevented it from
becoming known as the `M-play'. M-theory acquired its name from a similar
aversion, in this case of string theorists to the word `Membrane'. The
{\sl story of M} is thus the story of membranes, supermembranes in
particular. The occasion of Stephen Hawking's 60'th birthday is an
appropriate one for me to put on record some recollections of this story
because Stephen was exceptional in giving his support and encouragement
to work on supermembranes during the years in which the `M-word' could not
be pronounced. Thank you, Stephen, and happy 60th birthday.

A membrane is of course just a special case of a brane and, as the
reader will probably know, M-theory is really an Orwellian
democracy in which there are many equal branes but with some 
being more equal than others.  Strings are more equal for all the usual
reasons, but membranes are more equal too, for a different set of reasons.
In the light-front gauge, membranes are equivalent to the large $n$ limit
of $SU(n)$ gauge theories, dimensionally reduced to a quantum mechanical
model. The M(atrix) model formulation of M-theory could have been, and
nearly was, found from the 11-dimensional supermembrane in this way. But
this is all well-known, and given my spacetime limitations I prefer to
reminisce on the (pre)history of the 11-dimensional supermembrane.

This will be a selective history, chosen to motivate  discussion of a
surprising, and little-known, fact\footnote{In the talk I
explained how a tubular but axially {\sl asymmetric} supermembrane,
supported against collapse by angular momentum, can be both
stable and supersymmetry-preserving \cite{MNT1}. This
did surprise some members of the audience, although I discovered that the
stability issue had been previously adressed in a non-supersymmetric
context \cite{carter}. As a full discussion is available in
\cite{MNT1} and subsequent papers \cite{MNT2,KMPW}, I have chosen
to discuss another surprising fact about supermembranes in this
write-up.}: the field theory limit of a supermembrane in certain
hyper-K\"ahler backgrounds is a 3-dimensional sigma-model
with $N=4$ supersymmetry, but the supermembrane itself generically has
only $N=3$ supersymmetry \cite{GGPT}. This is a sigma-model analogue of the
breaking of $N=4$ to $N=3$ supersymmetry in 3-dimensional gauge theories
by the addition of a Chern-Simons term \cite{Lee}.

\section{The supermembrane}

It is well known that string theory arose from attempts to understand
the physics of hadrons. What is less well-known is that M-theory
has roots in hadron physics too. In 1978, the same year that
11-dimensional supergravity appeared \cite{CJS}, a `classical' bag model
for hadrons was proposed by Aurilia, Christodoulou and Legovini
\cite{ACL}; this was based on the idea that that the closed QCD 4-form
$Tr (F \wedge F)$ should be replaced, in an effective description of
hadrons, by an abelian 4-form field strength $G=dC$. Hadrons were
identified as those regions in which $G$ acquires a non-zero expectation
value; these regions would be  separated from the vacuum by a membrane
coupled to the 3-form potential $C$. I heard about this model from
Antonio Aurilia in 1980 and realized that the 4-form field strength
$G$ of 11-dimensional
supergravity could be similarly used, after reduction on $T^7$, to
introduce a positive cosmological constant into N=8  D=4 supergravity.
This supergravity theory had recently been constructed
by Cremmer and Julia \cite{CJ} but they had eliminated
the surviving four-dimensional 4-form field strength as if it were a
non-dynamical auxiliary field.  If one instead uses the field equation
of the 3-form potential $C$ then a positive cosmological constant appears
as the square of an integration constant\footnote{This idea occurred
independently to Duff and Van Nieuwenhuizen \cite{DvN} but
without the connection to 11-dimensional supergravity.}. We enlisted
Hermann Nicolai to help construct the new N=8 supergravity theory, which
turns out to have a positive scalar potential rather than a cosmological
constant \cite{ANT}. As this potential has no critical points it was
unclear what use it might have\footnote{Stephen Hawking used the idea of a
dynamical cosmological constant in his suggestion that the
`cosmological constant is probably zero' \cite{SWH} but it now seems
that it probably isn't zero.}. We should have continued this research by
considering whether a non-vanishing 4-form in D=4 could be combined with
compactifications on spaces other than $T^7$. We didn't, but Freund and
Rubin did \cite{FR} and their demonstration that D=11 supergravity could
be compactified on a 7-sphere sparked off the revival of interest in
Kaluza-Klein theory.

It was somehow forgotten, in all the Kaluza-Klein excitement, that the
3-form potential $C$ could couple to membranes (although
Bernard Julia was aware of the possibility \cite{BJ}). I think that the
main reason for this collective amnesia was the fact that 11-dimensional
supergravity was being promoted as a candidate unified field theory, so
the apparent absence of anything to which it could couple was viewed as an
advantage. This attitude discouraged thinking about membranes, which
didn't resurface until after the superstring revolution of 1984.
Following the construction by Green and Schwarz of a
covariant superstring action \cite{GS}, it was natural to reconsider the
possibility of an 11-dimensional supermembrane. During the summer of
1986, Luca Mezincescu and I attempted to construct a supermembrane
generalization of the Green-Schwarz (GS) action but the attempt 
did not succeed because we
were unable to generalize the self-dual worldsheet vector parameter of
the GS `$\kappa$-symmetry'; this made the two-dimensionality of the
string worldsheet seem an essential feature of the construction.
In fact, it is not; there is an alternative, but equivalent, form of the
$\kappa$-symmetry transformation with a worldsheet {\sl scalar} parameter.
This was found by Hughes, Liu and Polchinski in their construction of an
action for a super-3-brane in a D=6 Minkowski background \cite{HLP}; they
were motivated by the observation that a vortex of the D=6 supersymmetric
abelian-Higgs model is a supersymmetry-preserving 3-brane for which the
effective action must be of GS-type. I saw this paper the day before I was
to travel to Trieste to continue a collaboration with Eric Bergshoeff and
Ergin Sezgin, and soon after my arrival we suceeded in constructing an
11-dimensional supermembrane action that is consistent in any background
that solves the field equations of 11-dimensional supergravity
\cite{BST}. For unit tension the action takes the form
\be\label{bst}
S= - \int \left[{\rm Vol} \ \pm \ {\cal C}\right]
\ee
where ${\rm Vol}$ is the (appropriately defined)
induced volume 3-form, and ${\cal C}$
is the worldvolume 3-form induced by the {\sl superspace} 3-form
potential of 11-dimensional supergravity (of which $C$ is the bosonic
truncation). The choice of relative sign corresponds to the
choice between a supermembrane and an anti-supermembrane, or an $M2$-brane
and an ${\overline {M2}}$-brane in modern terminology.

A feature of all GS-type super-brane actions is that the fermions are
(apparently) worldvolume scalars. If this were really true then, for
example, the GS superstring action could not be equivalent to the worldsheet
supersymmetric NSR superstring action. In fact, the GS fermions are {\sl
not}
scalars because they are subject to the $\kappa$-symmetry gauge
transformation; it is for a similar reason that the 4-vector
potential of electrodynamics is not really a 4-vector field. To determine
the
transformation properties of the GS fermions under any symmetry of the
action (which would include spacetime Lorentz transformations for a
Minkowski background) one must first fix the
$\kappa$-symmetry gauge; the transformation is then a superposition of
the `naive' transformation with whatever compensating
$\kappa$-symmetry transformation is needed to maintain the gauge choice.
The gauge fixing must break the spacetime Lorentz group but can be chosen
to preserve the worldvolume Lorentz subgroup, under which the
gauge-fixed GS fermions turn out to transform as worldvolume spinors.

This transformation from spacetime spinor to worldvolume spinor is clearly
necessary if any spacetime supersymmetries are to be interpreted as
worldvolume supersymmetries after gauge-fixing, but it is not obviously
sufficient. In fact, initially it was far from clear that spacetime
supersymmetry would imply worldvolume supersymmetry of the
supermembrane, partly because the supermembrane has no NSR
formulation, and Ach\'ucarro, Gauntlett, Itoh and I
went to great lengths to verify it directly
\cite{AGIT}; our article was originally entitled {\sl Supersymmetry on the
brane} but we had to change the title to accomodate a referee who insisted
that use of the word `brane' would bring the physics community into
disrepute\footnote{Possibly this referee had in mind the earlier use of the
word in the 1954 essay {\sl Akquire culture and keep the brane clean} by
Nigel Molesworth
\cite{molesworth}. As this essay's subtitle is {\sl How to be Topp in Latin}
it is regretable that it fails to provide the modern translation of {\it
mens sana in corpore sano} which is {\it clean brane in clean bulk},
otherwise known as the braneworld cosmological principle.}.

Nowdays, the connection between spacetime supersymmetry and worldvolume
supersymmetry is considered obvious. However, as I hope the following
discussion will show, surprises are still possible.

\section{Backgrounds of reduced holonomy}

The $G\wedge G\wedge C$ term of 11-dimensional supergravity preserves
spacetime parity if $C$ is taken to be parity-odd, and with this parity
assignment the coupling of $G$ to fermion bilinears also preserves parity
because of the peculiar way that fermion bilinears behave under parity in
odd dimensions \cite{GN}. Thus 11-dimensional supergravity preserves
parity. It follows that solutions breaking parity must come in parity
doublets, each of which will preserve the same fraction of supersymmetry
because parity commutes with supersymmetry. We shall be interested in
solutions with vanishing $G$, and product 11-metric of the form
\be
ds^2_{11} = ds^2(\bE^{(1,2)}) + g_{IJ}(X) dX^I dX^J
\ee
where $g_{IJ}$ ($I,J =1,\dots,8$) is the metric of some Ricci-flat
8-dimensional manifold ${\cal M}_8$, or its orientation reversal 
$\overline {\cal M}_8$. 
Any submanifold with fixed position on ${\cal M}_8$ is a minimal surface
that we
may identify as the Minkowski vacuum of an infinite planar supermembrane.
In the gauge in which the worldvolume coordinates $\xi^i$ are identified
with
coordinates for $\bE^{(1,2)}$, the physical bosonic worldvolume fields of
the $M2$-brane are maps $X^I(\xi)$ from the worldvolume to ${\cal M}_8$, and
the bosonic action is
\be\label{SM}
I = -\int d^3\xi \, \sqrt{-\det (\eta_{ij} + g_{ij})}
\ee
where $\eta$ is the 2+1 Minkowski metric and
\be
g_{ij}(\xi) = \partial_i X^I \partial_j X^J g_{IJ}(X)\, .
\ee

To incorporate the fermions one may begin by noting that the gauge choice
breaks the 11-dimensional Lorentz group to the product of the
3-dimensional Lorentz group $Sl(2;\bR)$ with $SO(8)$. A 32-component
spinor of $SO(1,10)$ decomposes into the sum of the $({\bf 2},{\bf
8}_s)$ and $({\bf 2},{\bf 8}_c)$ irreps of this product group, where
${\bf 8}_s$ is the spinor representation of $SO(8)$ and ${\bf 8}_c$ is
the conjugate spinor representation. Only one of these two irreps
of $Sl(2;\bR)\times SO(8)$ survives the $\kappa$-symmetry gauge-fixing,
which one depending on whether the covariant action is the one for the
$M2$-brane or the one for the ${\overline {M2}}$-brane. By convention, we
shall take the physical fermion fields of the $M2$-brane to be in the
$({\bf 2},{\bf 8}_c)$ representation and those of the ${\overline {M2}}$-brane
to be in the $({\bf 2},{\bf 8}_s)$ representation. The spacetime parity
transformation that interchanges ${\cal M}_8$ with 
$\overline {\cal M}_8$ will also
interchange the ${\bf 8}_s$ and ${\bf 8}_c$ representations of $SO(8)$,
and hence will interchange $M2$ with ${\overline {M2}}$. Thus, an $M2$-brane
in  $\bE^{(1,2)} \times {\cal M}_8$ is equivalent to an
${\overline {M2}}$-brane in $\bE^{(1,2)} \times \overline{\cal M}_8$. In the
case that ${\cal M}_8$ has an orientation reversing isometry we have ${\cal
M}_8 \cong \overline{\cal M}_8$ and the $M2$-brane action will be equivalent to
the ${\overline {M2}}$-brane action.

Note that the field content of the gauge-fixed supermembrane is
bose-fermi balanced, as would be required for worldvolume supersymmety.
Whether the supermembrane {\sl is} worldvolume supersymmetric will depend
on the choice of ${\cal M}_8$. This follows  from the fact that
(super)symmetries of the supermembrane action arise from (super)isometries
of the background that leave invariant the superspace 4-form field strength.
In particular, for bosonic backgrounds of the type under consideration,
supersymmetries arise from Killing superfields whose spinor component is
a Killing spinor of M, and these exist only if ${\cal M}_8$ has special
holonomy. 

Let $H \subset SO(8)$ be the holonomy group. The number $N$ of
linearly-realized supersymmetries of the supermembrane is the number of
singlets in the decomposition of the spinor representation
${\bf 8}_s$ of $SO(8)$ into irreps of $H$. The number $N'$ of {\sl
non-linearly} realized supersymmetries is the number of singlets of the
${\bf 8}_c$ representation of $SO(8)$ in its decomposition into irreps of
$H$. For the anti-supermembrane the numbers $N$ and $N'$ are interchanged.
The groups $H$ for which $N>0$ fall into one of two nested sequences.
One sequence is
\be
G_2 \supset SU(3) \supset SU(2).
\ee 
The corresponding types of 8-manifold, and the values of $N$ and $N'$
are given in Table 1 (where $CY_n$ is a $2n$-dimensional Calabi-Yau
$n$-fold and $HK_{4n}$ is a hyper-K\"ahler manifold of quaternionic
dimension $n$).

\begin{table}[h]
\centering
\caption{}
\begin{tabular}{|c|c|c|c|}
\hline
H & ${\cal M}_8$ & $N$ & $N'$\\
\hline\hline
$G_2$ & $M_7 \times \bE^1$ & 1 & 1\\
\hline
$SU(3)$ & $CY_3\times \bE^2$ & 2 & 2\\
\hline
$SU(2)$ & $HK_4 \times \bE^4$ & 4 & 4 \\
\hline
\end{tabular}
\end{table}

In each of these cases the 8-manifold ${\cal M}_8$ takes the form
${\cal M}_8= {\cal M}_{8-k}
\times \bR^k$ ($k=1,2,4$) for some irreducible (8-k)-dimensional
manifold ${\cal M}_{8-k}$. Such 8-manifolds have an 
orientation-reversing isometry,
so the $M2$-brane in these backgrounds is equivalent to the ${\overline
{M2}}$-brane. 
In fact, they are identical because an anti-membrane can be obtained from
a membrane by a rotation in some $\bE^3$ subspace of $\bE^{2+k}$.
The reason that the $M2$ and ${\overline {M2}}$ actions can be identical is
that
their $\kappa$-symmetry transformations differ and this difference can
compensate for the different sign in (\ref{bst}).

Note that fixing the
position in ${\cal M}_{8-k}$ yields a supermembrane  in a Minkowski
spacetime
of dimension $D=4,5$ or $7$, according to whether $k=1,2$ or $4$,
respectively; as it happens, these are precisely the other dimensions for
which the supermembrane action is classically consistent \cite{BST}, so
the existence of these lower-dimensional supermembrane actions is 
explained by the existence of the 11-dimensional supermembrane.

The other sequence of holonomy groups is
\be
Spin(7) \supset SU(4) \supset Sp_2 \supset Sp_1\times Sp_1.
\ee
The corresponding types of 8-manifold, and the values of $N$ and $N'$
are given in Table 2.
\begin{table}[h]
\centering
\caption{}
\begin{tabular}{|c|c|c|c|}
\hline
H & ${\cal M}_8$ & $N$ & $N'$\\
\hline\hline
$Spin(7)$ & $Spin(7)$ & 1 & 0\\
\hline
SU(4) & $CY_4$ & 2 & 0\\
\hline
$Sp_2$ & $HK_8$  & 3 & 0 \\
\hline
$Sp_1\times Sp_1$ & $HK_4 \times HK_4$ &4&0\\
\hline
\end{tabular}
\end{table}
In each of these cases there are no non-linearly-realized supersymmetries,
so replacing the $M2$-brane by the ${\overline {M2}}$-brane breaks all
supersymmetries. Equivalently, replacing the 8-manifold ${\cal M}_8$ by its
orientation reversal $\overline {\cal M}_8$ breaks 
all $N$ supersymmetries of the
$M2$-brane action\footnote{Note that the background solutions preserve
$N$ supersymmetries irrespective of the orientation of ${\cal M}_8$; it  is
the only the rigid worldvolume supersymmetries on the 
supermembrane that are broken
when ${\cal M}_8$ is replaced by $\overline {\cal M}_8$. This is in contrast to
the related phenomenon of supergravity solutions with non-zero $G$ that are
supersymmetric for one orientation but non-supersymmetric for the other
orientation \cite{DNP}.}.

\section{The sigma model limit}

The action (\ref{SM}) can be expanded as a power series in $\partial X$.
Discarding a constant and terms with more than two derivatives we arrive at
the field theory action
\be
S = -{1\over2} \int d^3\xi \sqrt{-\det \eta}\, \eta^{ij} \partial_i X^I
\partial_j X^J g_{IJ}(X)\, .
\ee
This is a D=3 sigma-model with the 8-manifold ${\cal M}_8$ as its target
space. If the supermembrane preserved $N$ supersymmetries then an
analogous expansion yields a supersymmetric sigma-model with at least
$N$ supersymmetries. In most cases one can easily see that it can have no
more than $N$ supersymmetries because of the constraints imposed on the
target space of a sigma-model by extended supersymmetry; specifically, a
supersymmetric D=3 sigma model with an irreducible target space has $N=2$
supersymmetry if the target space is K\"ahler and $N=4$ if it is
hyper-K\"ahler \cite{sigma}. For example, because $Spin(7)$ manifolds are
not K\"ahler we know that the sigma-model obtained from the supermembrane
action can have at most $N=1$ supersymmetry. The same is true for the $G_2$
case, although the conclusion is less immediate in this case because the
8-manifold is not irreducible. Note that for the $N=2$ case of either
Table the target space is K\"ahler, as consistency requires,
but not hyper-K\"ahler, so the sigma-model has $N=2$  supersymmetry.
Similarly, for both $N=4$ cases the target space is hyper-K\"ahler, as
required for consistency. This leaves only the case of $Sp_2$ holonomy
of Table 2 to
consider, and here we find a surprise. As one sees from Table 2, the
gauge-fixed supermembrane action has only $N=3$ supersymmetry but, as its
target space is hyper-K\"ahler, the sigma model obtained from the field
theory limit has $N=4$ supersymmetry.  Thus, in this one case, the
low-energy sigma-model has {\it more} supersymmetries than the supermembrane
action from which it was derived!

From the sigma-model perspective, the supermembrane just adds
higher-dimension terms to the action. An interaction term that 
breaks $N=4$ supersymmetry to $N=3$ must also break
worldvolume parity. Majorana mass terms break parity in three dimensions
\cite{DJT} and although the supermembrane has no mass terms it does have a
mass parameter, determined by the membrane tension. Higher dimension
fermion interactions in the supermembrane must involve this parameter 
and so may break parity. For example, if $\psi$
is the 8-plet of real two-component $Sl(2;\bR)$ spinor fields then a term of
the form
\be
(\bar \psi \psi)(\bar \psi \gamma \cdot \partial \psi)
\ee
breaks parity for the same reason that Majorana mass terms break
parity. The supersymmetric completion of this term will not include any
purely bosonic term, consistent with parity preservation of the
bosonic truncation of the supermembrane, and it will not survive in the
field theory limit, consistent with parity preservation of the
supersymmetric sigma model.  Moreover, it can occur only when there are no
non-linearly realized supersymmetries, and hence must be absent in the cases
of Table 1.  Thus, a term of the above type is a candidate for a
parity-violating interaction that will break $N=4$ to $N=3$
supersymmetry when the hyper-K\"ahler target space has $Sp_2$ holonomy,
although it must be absent if the holonomy is contained in the $Sp_1\times
Sp_1$ subgroup of $Sp_2$.

There is a gauge theory precedent for all this. The addition of a
Chern-Simons to a 3-dimensional gauge theory with $N=4$ supersymmetry can
preserve at most $N=3$ supersymmetry, in which case its supersymmetric
completion will include parity-violating fermion mass terms \cite{Lee}.
In fact, this phenomenon is an M-theory dual of the one discussed here,
at least for the class of toric hyper-K\"ahler 8-manifolds, because
the M2-brane in such a background is dual to a D3-brane suspended between
$(p,q)$-fivebranes and the effective field theory on the
intersection is precisely a 3-dimensional $N=3$ gauge theory with
a Chern-Simons term \cite{GGPT,ohta}.

\medskip
\section*{Acknowledgments}
\noindent
I thank Jerome Gauntlett, Gary Gibbons, Paul Howe, Chris Pope and 
Kellogg Stelle for helpful discussions.


\begin{thebibliography}{9}



\bibitem{MNT1}
D. Mateos, S. Ng and P.K. Townsend, {\sl Tachyons, supertubes and
brane-antibrane systems}, JHEP {\bf 0203} (2002) 016.

\bibitem{carter}
B. Carter and X. Martin, {\sl Dynamic instability criterion for
circular string loops}, Ann. Phys (N.Y.) {\bf 227} (1993) 151.

\bibitem{MNT2}
D. Mateos, S. Ng and P.K. Townsend, {\sl Supercurves}, hep-th/0204062.

\bibitem{KMPW}
M. Kruczenski, R.C. Myers, A.W. Peet and D.J. Winters, {\sl Aspects of
supertubes}, hep-th/0204103.

\bibitem{GGPT}
J.P. Gauntlett, G.W. Gibbons, G. Papadopoulos and P.K. Townsend,
{\sl Hyper-K{\"a}hler manifolds and multiply intersecting branes},
Nucl. Phys. {\bf B500} (1997) 133.

\bibitem{Lee}
H-C. Kao and K. Lee, {\sl Self-dual Chern-Simons Higgs systems with
$N=3$ extended supersymmetry}, Phys. Rev. {\bf D46} (1992) 4691.

\bibitem{CJS}
E. Cremmer, B. Julia and J. Scherk, {\sl Supergravity theory in eleven
dimensions}, Phys. Lett. {\bf 76B} (1978) 409.

\bibitem{ACL}
A. Aurilia, D. Christodoulou and F. Legovini, {\sl A classical
interpretation of the bag model for hadrons}, Phys. Lett. {\bf B73}
(1978) 429. 

\bibitem{CJ}
E. Cremmer and B. Julia, {\sl The $SO(8)$ supergravity},
Nucl. Phys. {\bf B159} (1970) 141.

\bibitem{DvN}
M.J. Duff and P. Van Nieuwenhuizen, {\sl Quantum inequivalence of
different field representations}, Phys. Lett. {\bf B94} (1980) 179.

\bibitem{ANT}
A. Aurilia, H. Nicolai and P.K. Townsend, {\sl Hidden constants: the
theta parameter of QCD and the cosmological constant of N=8
supergravity}, Nucl. Phys. {\bf B176} (1980) 509.

\bibitem{SWH}
S.W. Hawking, {\sl The cosmological constant is probably zero},
Phys. Lett. {\bf B134} (1984) 403.

\bibitem{FR}
P.G.O. Freund and M.A. Rubin, {\sl Dynamics of dimensional reduction},
Phys. Lett. {\bf B97}, (1980) 233.

\bibitem{BJ}
B. Julia, {\sl Extra dimensions: recent progress using old ideas},
in  proceedings of the 2nd Marcel Grossman Meeting, Trieste, Italy,
1979. 

\bibitem{GS}
M.B. Green and J.H. Schwarz, {\sl Covariant description of
superstrings}, Phys. Lett. {\bf B136}, (1984) 367.

\bibitem{HLP}
J. Hughes, J. Liu and J. Polchinski {\sl Supermembranes},
Phys. Lett. {\bf B180} (1986) 370.

\bibitem{BST}
E. Bergshoeff, E. Sezgin and P.K. Townsend, {\sl Supermembranes and
11-dimensional supergravity}, Phys. Lett. {\bf B189} (1987) 75.

\bibitem{AGIT}
A. Ach{\' u}carro, J. Gauntlett, K. Itoh and P.K. Townsend,  {\sl
Worldvolume supersymmetry from spacetime supersymmetry of the four
dimensional supermembrane}, Nucl. Phys. {\bf B314} (1989) 129.

\bibitem{molesworth}
G. Williams and R. Searle, {\sl How to be Topp}, Max Parish \& Co.,
1954; republished in {\it Molesworth}, Penguin 1999. 

\bibitem{GN}
M.B. Gavela and R.I. Nepomechie, {\sl Discrete symmetries in
Kaluza-Klein theories}, Class. Quant. Grav. {\bf 1} (1984) L21.

\bibitem{DNP}
M.J. Duff, B.E.W. Nilsson and C.N. Pope, {\sl Spontaneous symmetry
breaking by the squashed seven-sphere}, Phys. Rev. Lett. {\bf 50}
(1983) 2043. 

\bibitem{sigma}
L. Alvarez-Gaum\'e and D.Z. Freedman, {\sl Geometrical structure and
ultraviolet finiteness in the supersymmetric sigma model},
Commun. Math. Phys. {\bf 80} (1981) 443.

\bibitem{DJT}
S. Deser, R. Jackiw and S. Templeton, {\sl Topologically massive gauge
theories}, Ann. Phys. (N.Y.) {\bf 140} (1982) 372.

\bibitem{ohta}
T. Kitao and N. Ohta, {\sl Spectrum of Maxwell-Chern-Simons theory realized
on type IIB brane configurations}, Nucl. Phys. {\bf B578} (2000) 215.


\end{thebibliography}
\end{document}